\documentclass[article]{revtex4}
\usepackage{graphicx}
\usepackage{amssymb}
\usepackage{bm}
\usepackage{hyperref}
\usepackage{epstopdf}
\usepackage{subfigure}

\begin{document}

\title{SNe Data Analysis in Variable Speed of Light Cosmologies  without Cosmological Constant}

\author{Pengfei Zhang$^{1}$\email{wjzpf@mail.nankai.edu.cn}, and Xinhe Meng$^{1,2}$\email{xhm@nankai.edu.cn}}

\affiliation{
$^1$Department of Physics, Nankai University, Tianjin 300071, China\\
$^2$State Key Laboratory of Theoretical Physics China,CAS, Beijing 100190, China\\Kavli Institute of Theoretical Physics China,CAS, Beijing 100190, China\\
{xhm@nankai.edu.cn(correspondence}}

\date{\today}

\begin{abstract}
In this work, we aim to show the possibilities of the variable speed of light (VSL) theory in explaining the type Ia supernovae observations without introducing dark energy. The speed of light is  assumed to be scale factor dependent, which is the most popular assumption in VSL theory. We show the modified calculation of the distance modulus, and the validity of the redshift-scale factor relation in VSL theory. Three different models of VSL are tested SNe data-sets with proper constraints on the model parameters. The comparison of the three models and flat $\Lambda$CDM in distance modulus is showed. Some basic problems and the difficulties of the confirmation of the VSL theory are also discussed.
\end{abstract}

\maketitle

\section{Introduction}
The variabilities of the constants in physics have been long studied. The speed of light in the vacuum which is labelled $c$ is one of them. The property of the speed of light is widely studied and the understanding of the speed of light has greatly progressed. In cosmology, the "Standard Big Bang"(SBB) model of the universe is accepted by most of the physicists, but there are still some puzzles to be explained. Recently, the VSL theory had been proposed. Albrecht and Magueijo proposed a theory that introduced a time-variable $c$ which is faster in the early universe {\cite{1}}, and showed how can their VSL theory solve some cosmological puzzles in the standard big bang model such as the horizon problem, the flatness problem, the entropy problem, the cosmological constant problem and some other problems. Moffat in his paper proposed a VSL theory with the speed of light having a phase transition {\cite{2}} which is called bimetric theory and took it as an alternative to inflation. Einstein field equations in Friedmann-Robertson-Walker spacetime in the VSL theory had been discussed by Barrow{\cite{3}} and the geometrodynamics of the VSL cosmologies had been discussed by Bassett{\cite{4}}. In their papers they argued that the time-variable $c$ should not introduce changes in the Friedmann equations in the cosmological frame and Einstein equations must remain effective.
In 1999, the high-redshift observations of quasar absorption spectra showed the fine-structure constant $ \alpha=e^{2}/\hbar{c}$ varies with the cosmic time{\cite{5}}. The following research showed that the fine-structure $\alpha$ doesn't only rely on the cosmic time but also relies on the space {\cite{alpha1} \cite{alpha2}}. The physicists proposed many theories to explain the time-space dependent $\alpha$. Because of the fine-structure $\alpha$ is inverse proportional to the speed of light, the change of speed of light gives rise to the change of $\alpha$,  the VSL theory began to draw more attentions of the physicists. Though a mass of papers had been published and claimed that the VSL theory could solve a lot of puzzles, there are few papers to constrain the VSL models with the observation data.

Albrecht, Majueijo, Barrow and Bassett {\cite{1}\cite{3}\cite{4}} have shown that a time-variable c should not introduce changes in the curvature in Einstein's equations in the cosmological frame and the Einstein equation in VSL theory are still valid. By choosing the co-moving proper time to be the the specific choice of time coordinate, assuming our universe is spatially homogeneous and isotropic, leads to the requirement that the Friedmann equations still retain their form with $c(t)$ and $G(t)$ varying. The Einstein equations and Friedmann equations are given as follows:
\begin{equation}\label{1}
G_{\mu\nu}-g_{\mu\nu}\Lambda=\frac{8\pi G(t)}{c^{4}}T_{\mu\nu},
\end{equation}
\begin{equation}\label{2}
\frac{\dot{a}^{2}}{a^{2}}=\frac{8\pi G(t)\rho}{3}-\frac{K c(t)^{2}}{a^{2}},
\end{equation}

\begin{equation}\label{3}
\frac{\ddot{a}}{a}=-\frac{4\pi G(t)}{3}(\rho+\frac{3p}{c(t)^{2}}),
\end{equation}
where, $a$ is the scale factor, $\rho$ and $p$ are the density of the matter and the pressure of the matter respectively, $K$ is the metric curvature parameter.

In this paper, we use the Friedman-Lema\^{\i}tre-Robertson-Walker metric:
\begin{equation}\label{4}
ds^{2}=-c^{2}(t)dt^{2}+a^{2}(t)\left[ \frac{dr^{2}}{1-Kr^{2}}+r^{2}(d\theta^{2}+\sin ^{2}\theta {}d\phi ^{2})\right],
\end{equation}
and we will take the curvature parameter $K=0$ in following discussions.

In this paper, we aim to constrain some specific VSL models with the SNe data to investigate whether the models could fit the astrophysics observations. The expansion of the universe is thought to be accelerating by the recent observation of type Ia supernovae{\cite{sn1}}{\cite{sn2}}, and the acceleration has been attributed to a mysterious component dubbed dark energy. The cosmological constant $\Lambda$ is the simplest candidate for dark energy, and $\Lambda$CDM model appears to explain the astrophysical observations satisfactorily. Here, we do not introduce the dark energy (the cosmological constant $\Lambda$ is set to be zero ), the possibilities for the VSL theory to explain the dark energy are shown.

This paper is arranged as follows, in this section, we show some basic existing conclusions such as the Einstein equations and the Friedman equations under VSL assumptions given by Albrecht, Magueijo, Barrow and Bassett. In section II, the modified calculation of luminosity distance is shown, the validity of the redshift-scale factor relation is reexamined. In section III, three VSL models are studied. The results of data analysis are shown, and the best fitting value of the model parameters are given. The comparison of the three models and flat $\Lambda$CDM in distance modulus vs. redshift is shown.

\section{Some basic discussions}
In subsection A, we will discuss some fundamental issues on the measurement under VSL assumptions. The validity of the redshift-scale factor relation will be shown. In subsection B, we will give the modified luminosity distance with a time-varying speed of light.

\subsection{measurement under VSL assumptions}
Before we discuss the value of the speed of light, we must discuss how to measure the distance first. Because, in 1983, the unit "meter" has been defined as "the distance of the path travelled by light in vacuum during a time interval of $\frac{1}{299,792,458}$ of a second" by the International Bureau of Weights and Measures. Under this definition, the speed of light $c$ is defined to be 299,792,458 meters per second exactly. So, any discussion of the VSL theory before choosing a new definition of the unit "meter" is meaningless. Only after we measure the distance in a way that doesn't rely on the speed of light can we study the VSL theory. For example, Albrecht and Magueijo proposed to take the Bohr radius $r_{Bohr}=4\pi \epsilon_{0} \hbar^{2}/m_{e} e^{2}$ to be a standard distance that never change.
Then, we notice that almost all the modern astrophysical observations take the redshift $z$ as a basic observed quantity, which is defined as the fractional shift in wavelength of a photon, $z\equiv\frac{\lambda_{obs}-\lambda_{em}}{\lambda_{em}}$. And the relation $a=\frac{1}{1+z}$ is always considered effective,where, $a$ is the normalized scale factor when we set the today scale factor $a_{0}$ to be 1.

May the relation with redshift $z$ and scale factor $a$ change under VSL assumptions? In our paper, we will derive the redshift-scale factor relation before the data fitting. Now, let's consider an object that located in $r_{em}$ emits a photon at $t_{em}$ ,then the photon is observed by a detector located in original point at $t_{ob}$. Because the photon propagates along the null geodesic, we have:
\begin{equation}\label{5}
ds^{2}=-c(t)^{2}dt^{2}+a^{2}(t)[dr^{2}+r^{2}(d\theta^{2}+\sin ^{2}\theta {}d\phi ^{2})]=0.
\end{equation}

We choose $d\theta$ and $d\varphi$ to be zero, and that does not effect the conclusion, then we have
\begin{equation}\label{1.1}
\int_{t_{em}}^{t_{ob}}\frac{c(t)}{a(t)}dt=\int_{r_{em}}^{0}dr,
\end{equation}

\begin{equation}\label{1.2}
\int_{t_{em}+\Delta t_{em}}^{t_{ob}+\Delta t_{ob}}\frac{c(t)}{a(t)}dt= \int_{r_{em}}^{0}dr,
\end{equation}
where, $\Delta t_{em}$ and $\Delta t_{ob}$ are the time-interval when the photon is emitted and observed. Subtract the two equations (\ref{1.1}) and (\ref{1.2}), we have
\begin{equation}
\int_{t_{em}}^{t_{em}+\Delta t_{em}}\frac{c(t)}{a(t)}dt= \int_{t_{ob}}^{t_{ob}+\Delta t_{ob}}\frac{c(t)}{a(t)}dt,
\end{equation}
which shows
\begin{equation}\label{redshift 1}
\Delta t_{em}\times\frac{c(t_{em})}{a(t_{em})}=\Delta t_{ob}\times\frac{c(t_{ob})}{a(t_{ob})}.
\end{equation}
As the wavelength $\lambda$ of the photon is proportional to speed of light and time, we have
\begin{equation}\label{redshift 2}
\frac{\lambda_{ob}}{\lambda_{em}}=\frac{c_{ob}\Delta t_{ob}}{c_{em}\Delta t_{em}}.
\end{equation}
Combine equations (\ref{redshift 1}) and (\ref{redshift 2}), we will have the relation between redshift and the scale factor in VSL theory.
\begin{equation}
z\equiv\frac{\lambda_{ob}-\lambda_{em}}{\lambda_{em}}=\frac{a_{ob}}{a_{em}}-1.
\end{equation}
Now, we draw a conclusion that, the relation between redshift and scale factor is still valid. We can use the expression $a=\frac{1}{1+z}$ in our data analysis.

\subsection{The luminosity distance and distance modulus}
Consider a photon propagates along the null geodesic (set $d\theta$ and $d\phi$ to be $0$)
\begin{equation}\label{5}
ds^{2}=-c^{2}(t)dt^{2}+a^{2}(t)dr^{2}=0.
\end{equation}
We can rewrite the proper distance and the angular diameter distance:
\begin{equation}\label{6}
D_{h}=a_{0}r_{h}=a_{0}\int_{0}^{z}\frac{c(t)}{H(z')}dz',
\end{equation}
\begin{equation}\label{dadh}
D_{A}=a(t)r_{h}=a(t)\int_{0}^{z}\frac{c(t)}{H(z')}dz'=\frac{1}{1+z}D_{h},
\end{equation}
where, $r_{h}$ is the co-moving distance, $H(z)$ is the Hubble parameter.

The luminosity distance $D_{L}$ is introduced in order to link the supernova luminosity with the expansion rate of the Universe and it is defined by
\begin{equation}\label{L1}
D_{L}^{2}\equiv\frac{L_{s}}{4\pi\cal{F}},
\end{equation}
where $L_{s}$ is the absolute luminosity of a source and $\cal{F}$ is an observed flux. If we were in a static space the radiation observed flux would simply be $\mathcal{F}=\frac{L_{s}}{4\pi D_{h}^{2}}$, but the Universe is actually expanding and that affects the photons as they propagate from the source to the observer. There are actually two effects,
\begin{itemize}
\item The individual photons lose energy $\propto(1+z)$ because of the expansion of the Universe,  so they have less energy when they arrive.
\item The photons arrive less frequently $\propto(1+z)$.
\end{itemize}
Combining the two, the observed flux is
\begin{equation}\label{L2}
\mathcal{F} =\frac{L_{s}}{4\pi D_{h}^{2}(1+z)^{2}}.
\end{equation}
Combining equation~(\ref{L1}), equation~(\ref{dadh}) and equation~(\ref{L2}), we can obtain that the luminosity distance:
\begin{equation}\label{dl1}
D_{L}=(1+z)D_{h}=(1+z)^{2}D_{A}=(1+z)\int_{0}^{z}\frac{c(t)}{H(z')}dz'.
\end{equation}
For simplicity, we define the dimensionless Hubble parameter:
\begin{equation}\label{dhp}
E^{2}(z)\equiv\frac{H^{2}(z)}{H_{0}^{2}},
\end{equation}
$H_{0}$ is the Hubble constant.
Combine equation~(\ref{dhp}) and equation~(\ref{dl1}), we can rewrite the luminosity distance:
\begin{equation}\label{dl2}
D_{L}=(1+z)D_{h}=\frac{1+z}{H_{0}}\int_{0}^{z}\frac{c(t)}{E(z')}dz'.
\end{equation}
The distance modulus is given as follow
\begin{equation}\label{dm}
\mu(z)=5\log_{10}D_{L}(z)+\mu_{0},
\end{equation}
where $\mu_{0}$ have to be determined for best fitting to observational data.

\section{SNe data constraints}
In VSL theory, we take the speed of light to be a variable quantity. The speed of light is no more invariable but time-dependent or space-dependent. In this paper, the time-dependent $c$  is studied, we take the speed of light as a function of cosmic time. As the scale factor is the function of cosmic time, we can write the time-dependent $c$ to be a function of the scale factor, $c=c(a)$.

In this section, firstly we give the model that was proposed by Majueijo and Albercht \cite{1} and then propose two new phenomenological
 models $c=c_{0}e^{n(1-a)}$ and $c=c_{0}\frac{m}{n+a}$. In the following subsections , the three models of VSL are constrained by the SNe data ( Union 2.1~\cite{union}). In the studied models $H_{0}$ is a fixed parameter which we set to be $67.11 (km/s)/Mpc$ {\cite{Planck}} and that doesn't affect the fitting results, while $'n'$ is the free parameter in the models.

\subsection{Model I}
Majueijo and Albercht proposed a power law model of the speed of light:
 \begin{equation}\label{m1}
c(a)=c_{0}a^{n},
 \end{equation}
showing that $c$ varies with the scale factor, the model parameter $n$ is a constant, and $c_{0}$ is the modern speed of light. When $n$ is a negative number, we can see $c$ is vary large when $a$ is small and $c$ is infinite when $a=0$. Considering that we do not introduce the dark energy and the universe are constituted by non-relativistic matter (baryon component, dark matter) and the relativistic component. According to observations of the modern astrophysics, the relativistic component is negligible compared with the non-relativistic component. Thus, the expression of dimensionless Hubble parameter $E$ could be wrote as (take $\dot{G}=0$ ):
\begin{equation}\label{e}
 E=a^{-\frac{3}{2}}=(1+z)^{\frac{3}{2}},
\end{equation}

From equations (\ref{dm}) and (\ref{e}) we can obtain the specific expressions of distance modulus $\mu(z)$ of the power law model.
\begin{equation}
\mu(z)=5\lg(\int_{0}^{z}\frac{(1+z^{'})^{-n}}{H_{0}(1+z^{'})^\frac{3}{2}}dz^{'})+\mu_{0}=5\lg(\frac{1}{H_{0}}\int_{0}^{z}(1+z^{'})^{-n-\frac{3}{2}}dz^{'})+\mu_{0}
\end{equation}

Then we use the Bayesian analysis to constrain the model.
The fitting results of model I are shown in figure 1 and figure 2.
\begin{figure}[!h]
\centering
\includegraphics[scale=1, width=6.5cm, height=4.5cm]{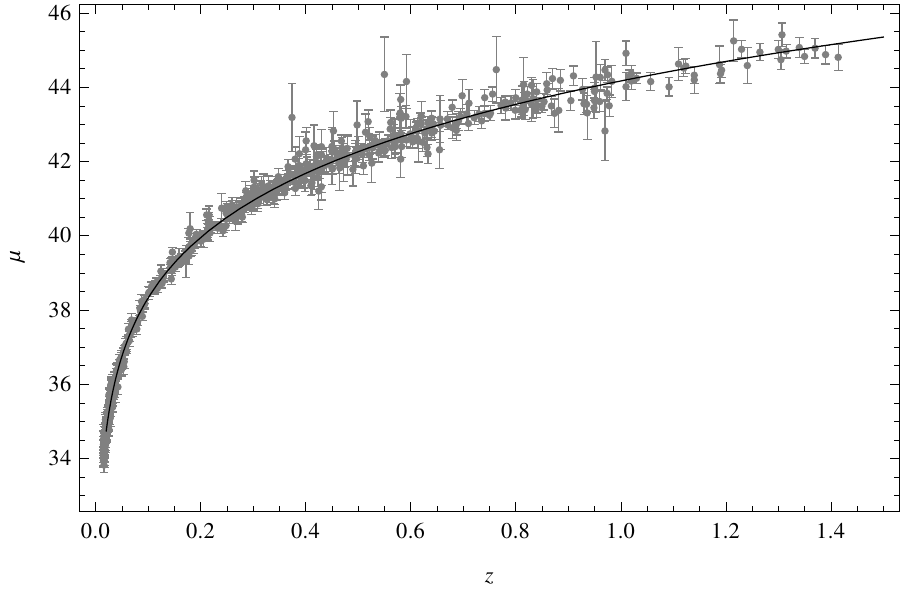}
\caption{The minimum of the $\chi^{2}$ is 568.313, when the parameters $\mu_{0}$ and $n$ take the value 34.05 and -0.861 respectively. The black curve is the best fitting line.}
\end{figure}

\begin{figure}[!h]
\centering
\includegraphics[scale=1, width=6.5cm, height=4.5cm]{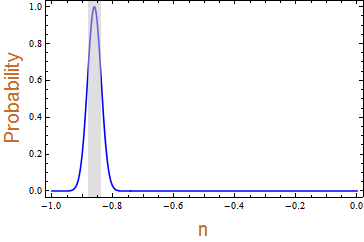}
\caption{The likelihood distribution of the model parameter n. The shaded region shows the $1\sigma$ limit. }
\end{figure}

We can see that the model I fits well the SNe data, and gives a minimum value of $\chi^{2}$  when the model parameter $n$ takes the value -0.861, which is consistent with our hypothesis that light speed is higher at higher redshift. In figure 2, we provide the likelihood distribution of the parameter $n$, and the shaded region which covers from $-0.886$ to $-0.828$ shows the $1\sigma$ limit for $n$.

\subsection{Model II}
In this subsection, we propose a exponential form function of $c$:
\begin{equation}\label{m2}
c(a)=c_{0}e^{n(1-a)},
\end{equation}
where, $n$ is a positive constant and $a$ is the normalized scale factor with $a_{0}=1$. The speed of light $c$ gradually decrease with the increase of $a$. In early universe the speed of light has a maximum, the limiting value of $c$ is $c_{max}=c_{0}e^{n}$ when $a$ takes the value 0. In this model, the speed of light does not have a quite large value in early universe, and the model does not have a singularity when $a=0$.

From equations (\ref{dm}), (\ref{e}) we can obtain the specific expressions of distance modulus $\mu(z)$ of the exponential form model:
\begin{equation}
\mu(z)=5\lg(\int_{0}^{z}\frac{e^{\frac{nz'}{1+z'}}}{H_{0}(1+z')^{3/2}}dz')+\mu_{0}
\end{equation}

Then we use the Bayesian analysis to constrain the model.
The fitting results of model II are shown in figure 3 and figure 4.
 \begin{figure}[!h]
\begin{center}
\includegraphics[scale=1, width=6.5cm, height=4.5cm]{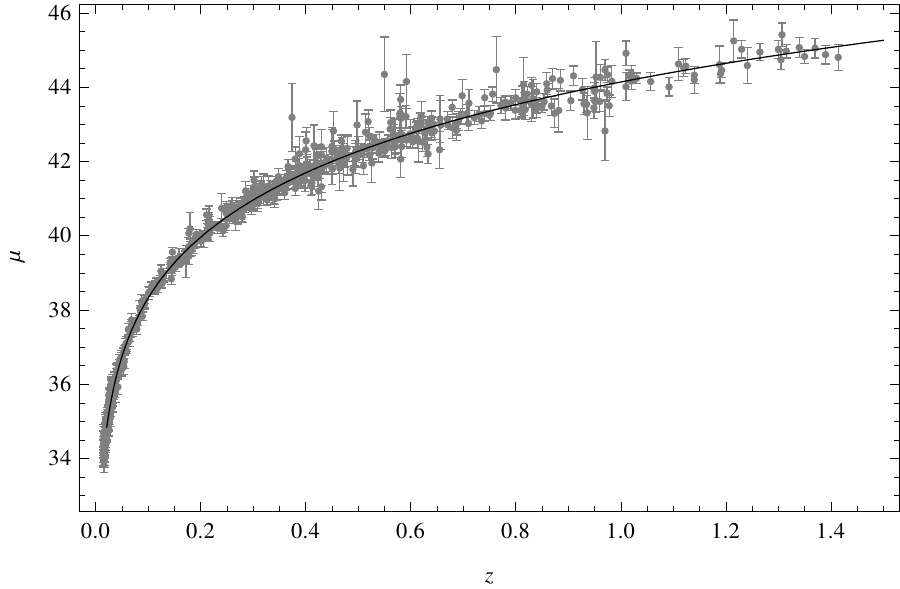}
\end{center}
\caption{The minimum of the $\chi^{2}$ is 562.609, when the parameters $n$ and $\mu_{0}$ take the value 1.0604, 34.032 respectively. The black curve is the best fitting line.}
\end{figure}
\begin{figure}[!h]
\centering
\includegraphics[scale=1, width=6.5cm, height=4.5cm]{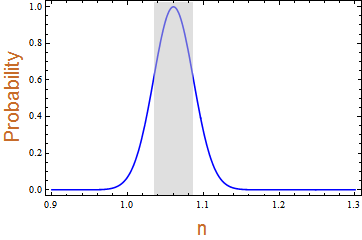}
\caption{The one-dimensional probability distribution function(PDF) for the model parameter n of model II (blue curve). The shaded region shows the $1\sigma$ limit.}
\end{figure}

From figure 3, we can see that the model II fits well the SNe data, and give a minimum value of $\chi^{2}$  when the parameter $n$ takes the value 1.06. In figure 4, we provide the likelihood distribution of the parameter $n$, and the shaded region which covers from $1.020$ to $1.099$ shows the $1\sigma$ limit for $n$. The result given by SNe data favours positive value of $n$, thus this model is also acceptable.

\subsection{Model III}
We propose another new model having a fractional form:
\begin{equation}\label{m3}
c(a)=c_{0}\frac{m}{n+a},
\end{equation}
the constants, $m$ and $n$, are model parameters, and $a$ is the scale factor.
The present speed of light is $c_{0}$, that will restrict the model parameters, $c_{0}=c_{0}\frac{m}{n+1}$. Then, $m=n+1$ must be stand. \\
The speed of light $c$ gradually decrease with the increase of $a$. In early universe, the speed of light $c$ has a maximum,  the limiting value is $c_{max}=c_{0}\frac{n+1}{n}$ when $a=0$.

From equations (\ref{dm}), (\ref{e}) we can obtain the specific expressions of distance modulus $\mu(z)$ of the fractional form model.
\begin{equation}
\mu(z)=5\lg(\int_{0}^{z}\frac{\frac{(1+n)(1+z')}{1+n+nz'}}{H_{0}(1+z')^{3/2}}dz')+\mu_{0}
\end{equation}

Then we use the $\chi^{2}$ analytical method to constrain the model.
The fitting results of model III are shown in figure 5 and figure 6.

 \begin{figure}[!h]
\begin{center}
\includegraphics[scale=1, width=6.5cm, height=4.5cm]{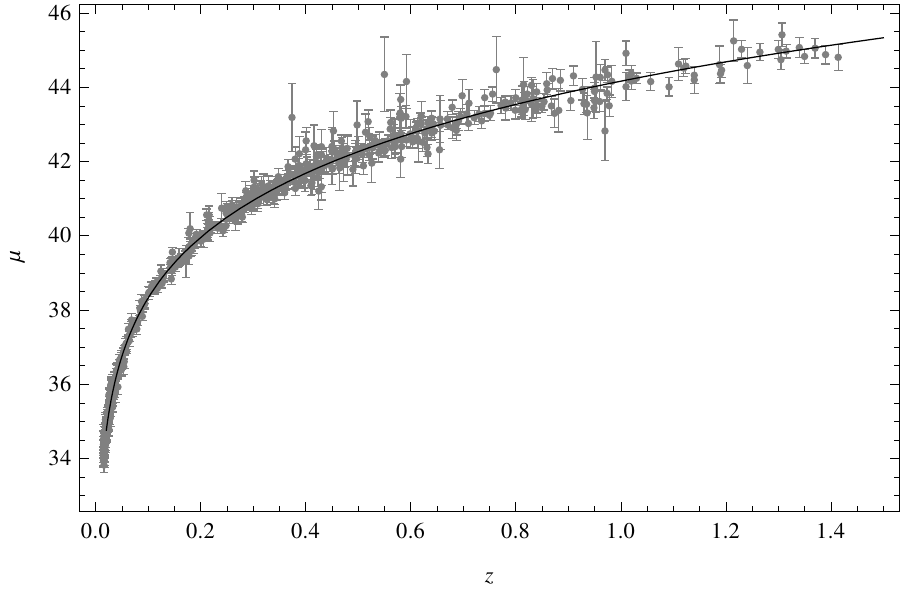}
\end{center}
\caption{The minimum of the $\chi^{2}$ is 566.968, when the parameters $n$ and $\mu_{0}$ take the value 0.1325, 34.048 respectively. The black curve is the best fitting line.}
\end{figure}

\begin{figure}[!h]
\centering
\includegraphics[scale=1, width=6.5cm, height=4.5cm]{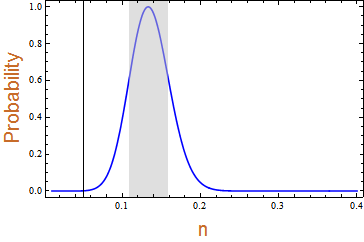}
\caption{The one-dimensional probability distribution function for the model parameter n of model III (blue curve). The shaded region shows the $1\sigma$ limit.}
\end{figure}

From figure 5, we can see that the model III fits well the SNe data, and give a minimum value of $\chi^{2}$  when the parameter $n$ takes the value 0.1325. And in figure 6, we provide the likelihood distribution of the parameter $n$, and the shaded region which covers from $0.096$ to $0.173$ shows the $1\sigma$ limit for $n$.

The best fitting results of the three model and the reduced $\chi^{2}$ are shown in table 1.

\begin{table}[!h]
\begin{center}
\caption{Best-fit parameters of VSL models by SNe, where the reduced $\chi^{2}$ is chi-square divided by the degree of freedom (which is $N-N_{m}-1$ ,where $N$ is the number of observations, and $N_{m}$ is the number of fitted parameters). }
\begin{tabular}{|c|c|c|c|c|c|c|}
  \hline
  Model & reduced $\chi^{2}_{min}$ & best fitting value of $n$ & 1$\sigma$ confidence interval of $n$ \\
  \hline
  Model I & 0.983 & -0.861 & (-0.8832, -0.8386)  \\
  Model II & 0.973 & 1.06  & (1.033, 1.086) \\
  Model III & 0.981 & 0.1325 & (0.1087, 0.1589)  \\
  \hline
\end{tabular}
\end{center}
\end{table}
\begin{figure}[!h]
\centering
\includegraphics[scale=1, width=6.5cm, height=4.5cm]{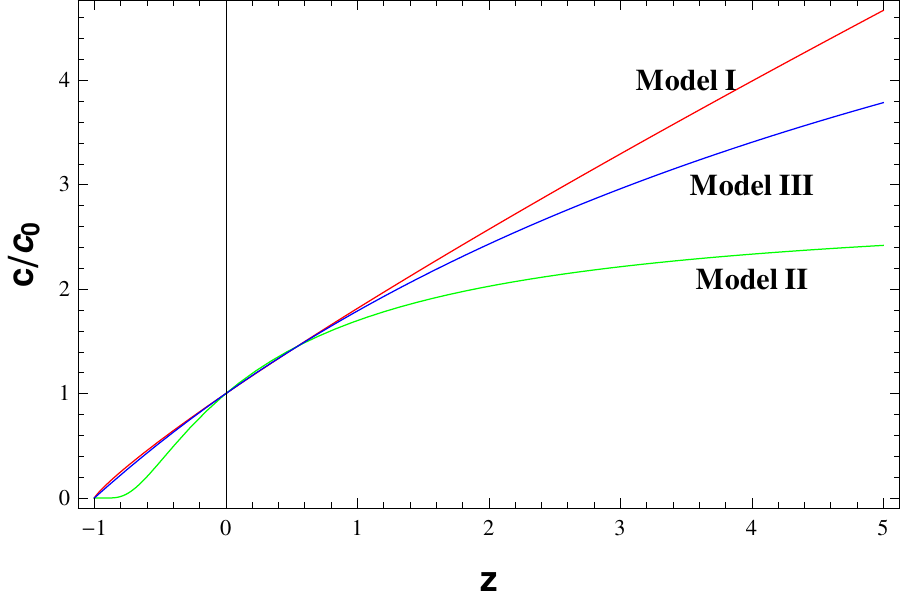}
\caption{The speed of light $c$ in different models when we restrain the model parameter to be the best fitting values. The red, green, blue curve for model I, II, III respectively.  }
\end{figure}

We set the model parameter $n$ to be the best fitting values. The speed of light $c$ in the three models are shown in figure 7. Form figure 7, we can see that, the speed of light $c$ in each models increase with the redshift $z$ and decrease with the scale factor $a$. In other words, the SNe data favor a gradually decreasing function of $c(a)$ in VSL theory.

In figure 8 and 9, we showed the comparison of the three models and flat $\Lambda$CDM in distance modulus vs. redshift. (In our fitting program, 0.72 and 34.025 is the best fitting value of $\Omega_{\Lambda}$ and $\mu_{0}$ in $\Lambda$CDM model). From figure 8, we can see that the distance moduli are quite similar in different models when $z$ ranges from 0 to 10. So, we showed the distance moduli when $z$ ranges frome 2 to 10 in order to demonstrate the differences. Maybe the four models couldn't be ruled out unless more data with larger redshift are used.

 \begin{figure}[!h]
\begin{center}
\includegraphics[scale=1, width=6.5cm, height=4.5cm]{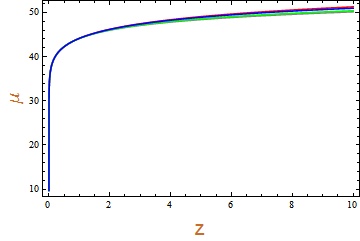}
\end{center}
\caption{The distance moduli vs. redshift in  model I, II, III and flat $\Lambda$CDM are given by the red, green, blue and gray curves respectively, here $z$ ranges from 0 to 10.  }
\end{figure}

 \begin{figure}[!h]
\begin{center}
\includegraphics[scale=1, width=6.5cm, height=4.5cm]{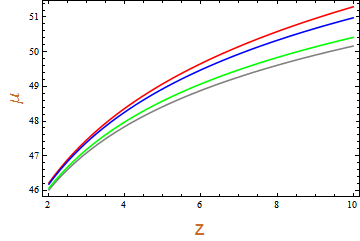}
\end{center}
\caption{The distance moduli vs. redshift in  model I, II, III and flat $\Lambda$CDM are given by the red, green, blue and gray curves respectively, here $z$ ranges from 2 to 10.  }
\end{figure}

\section{Conclusion and discussion }
Like other physical constants , the property of the speed of light and its variability have been long studied. The proposed VSL models are claimed to be capable of explaining many cosmological puzzles --- which exist in the "Standard Big Bang" universe --- by introducing a faster speed of light. The VSL theory is taken as the candidate of inflation theory, but the validity of the VSL theory according to the astrophysical observations are not well studied.

In this paper, the linear relation between redshift and the scale factor is proved to be valid under VSL assumptions, the new expression of the distance modulus is shown. We constrain three different VSL models with the SNe data, from the results in section III, we draw a conclusion that the VSL theory could fit the data well. We can see that the three time dependent VSL models can well fit the SNe data without introducing the cosmological constant $\Lambda$, the late time expansion of our universe could be explained without dark energy by introducing a faster speed of light.

Though the results of our data fitting given above could show the possibilities for the VSL theory to explain the astrophysical observations, many problems are not included, for example, we set the gravitational parameter $G$ to be constant in SNe data fitting. We just did a phenomenological analysis and more researches could be done before the validity of the VSL theory is confirmed. And similarly, though the vsl theory have been proposed for many years and lots of papers have been published, the VSL theory does not have a strong foundation and many problem are need to be solved or explained.
We will study more fundamental questions in further researches.

\section*{Acknowledgement}
We thanks the help from Xiaolong Du and Jiaxin Wang. This work is partly supported by National Natural Science Foundation of China under Grant Nos. 11075078 and 10675062 and by the project of knowledge Innovation Program (PKIP) of Chinese Academy of Sciences (CAS) under the grant No. KJCX2.YW.W10 through the KITPC where we have initiated this present work.

\end{document}